\documentstyle[12pt,epsfig]{article}

\columnwidth\textwidth

\def\slash#1{\rlap{$#1$}/ } 
\def\bigslash#1{\rlap{$#1$}\thinspace / \thinspace }
\def\slashD{{\bigslash D}}
\def\vD{{D \cdot v}}
\def\psibar{\overline{\psi}}
\def\si{\psi_{+v}}
\def\sim{\psi_{-v}}
\def\sibar{\overline{\psi}_{+v}}
\def\simbar{\overline{\psi}_{-v}}
\def\qint{\int {d^4q \over (2\pi)^4}}
\def\nl{\nonumber \\}

\begin{document}
\pagenumbering{arabic}
\everymath={\displaystyle}
\vspace*{-2mm}
\thispagestyle{empty}
\noindent
\hfill HUTP-96/A019\\
\mbox{}
\hfill hep-ph/9607272\\
\mbox{}
\hfill July 1996  \\
\begin{center}
  \begin{Large}
  \begin{bf} {\Large \sc 
Renormalizing the heavy quark effective field
theory
Lagrangian to order $1 \over m^2$}
   \\
  \end{bf}
  \end{Large}
  \vspace{0.8cm}
   Markus Finkemeier, 
   Matt McIrvin\\[2mm]
   {\em Lyman Laboratory of Physics\\
        Harvard University\\
        Cambridge, MA 02138, USA}\\[5mm]
{\bf Abstract} 
\end{center}
\begin{quotation}
\noindent
The heavy quark effective field theory Lagrangian is renormalized to
order $1 \over m^2$. Our technique eliminates operators that vanish by
the equation of motion by continuously redefining the heavy quark fields
during renormalization. It is consequently only necessary to calculate the
running of the operators that do not vanish by the equation of motion.
We show that our results are consistent with reparameterization
invariance.\footnote{
This is a long version of our paper, giving many details of the calculation.
A somewhat shortened version (HUTP-96/A026) is being submitted for
publication.}
\end{quotation}
%
%

%

\def\slash#1{\rlap{$#1$}/ } 
\def\bigslash#1{\rlap{$#1$}\thinspace / \thinspace }
\def\slashD{{\bigslash D}}
\def\vD{{D \cdot v}}
\def\psibar{\overline{\psi}}
\def\si{\psi_{+v}}
\def\sim{\psi_{-v}}
\def\sibar{\overline{\psi}_{+v}}
\def\simbar{\overline{\psi}_{-v}}
\def\qint{\int {d^4q \over (2\pi)^4}}
\def\nl{\nonumber \\}

\section{Tree-level matching and class II operators}

Heavy quark effective field theory (HQEFT) is an approximation to QCD
for quarks with masses that are large compared to the characteristic
momentum scale of strong interactions.  In the infinite-mass limit, it
possesses Isgur-Wise spin-flavor symmetry, which facilitates the
calculation of matrix elements of weak currents.  The corrections to the
theory for finite quark mass break Isgur-Wise symmetry, and these form
an infinite series in powers of the reciprocal of the quark mass.

There are two kinds of subleading operators which appear as terms in the
corrected Lagrangian. Some operators, known as class I operators, do not
vanish if the field is assumed to obey the leading-order classical
equation of motion; others, known as class II operators, do vanish under
this assumption.

Both types of operator appear in the Lagrangian obtained from matching
to QCD. Tree-level matching with full QCD to order $1\over m^2$
yields the Lagrangian
\begin{eqnarray}
        {\cal L} &=& \sibar i\vD \si
        + {1\over 2m} \sibar[(i\slashD)^2-(i\vD)^2]\si \nl
        &&+ {1\over 4m^2} \sibar[-i\slashD i\vD i\slashD + (i\vD)^3]\si
\end{eqnarray}
The leading-order equation of motion is
\begin{equation}
        D \cdot v \si = 0.
\end{equation}
Therefore, the operators
\begin{eqnarray}
        O_{kin} &=& -{1 \over 2m} \sibar D^2 \si \nonumber \\
        O_{mag} &=& {g \over 4m} \sibar \sigma^{\mu \nu} G_{\mu \nu} \si
                                \nonumber \\
        O_1     &=& {g \over 8m^2} \sibar v^{\mu} [D^{\nu}, G_{\mu \nu}]
                \si \nonumber \\
        O_2     &=& {ig \over 8m^2} \sibar \sigma^{\alpha \mu} v^{\nu}
                                \lbrace D_{\alpha},G_{\mu \nu}\rbrace \si
    \label{eq:operators}
\end{eqnarray}
are class I, whereas operators such as
\begin{equation}
        O_{\vD} = -{1 \over 2m} \sibar (i\vD)^2 \si
\end{equation}
are class II.  There is some freedom in defining what part of the
Lagrangian is class I and what part is class II, since the class I
operators may be defined to absorb part of the class II terms. Written
in terms of (\ref{eq:operators}), the Lagrangian is
\begin{eqnarray} \label{eq:l}
        {\cal L} &=& \sibar i\vD \si + O_{kin} + O_{mag}
        - {1\over 2m} \sibar (i\vD)^2 \si + O_1 + O_2 \nl
        &&+ {1\over 8m^2} \sibar [-(i\slashD)^2i\vD -i\vD(i\slashD)^2]\si
        + {1\over 4m^2} \sibar (i\vD)^3 \si
\end{eqnarray}

In many
treatments of HQEFT to order $1 \over m$, $O_{\vD}$ is simply thrown out
of the Lagrangian. It is not obvious, however, that simply throwing
out the class II operators is correct at 
higher order in $1/m$, and
therefore we will discuss this point is some detail.

The systematic method to remove redundant operators is via a suitable
field redefinition. 
In the case (\ref{eq:l}), the field redefinition
\begin{equation}
        \si \rightarrow \left[ 1 + {i \vD \over 4m}
        - {(i \vD)^2 \over 32m^2} \right] \si
        \label{eq:simpleredef}
\end{equation}
removes the class II operators;
in terms of the redefined
quark fields, it becomes ${\cal L}_0 + O_{kin} + O_{mag} + O_1 + O_2$.
Note that the field redefinition
(\ref{eq:simpleredef}) does not change the coefficients of the
class I operators to order $1 \over m^2$, and so in fact is equivalent
to the naive procedure of just dropping the class II terms. As we will
show now, this is {\em not} the case in general.

Class II operators can be removed from the effective Lagrangian 
to any desired order by use of the following iterative procedure.
For this procedure to be possible, the 
the class I and class II operators must be defined such that they are
{\it separately} Hermitian (note that this is not fulfilled by the operator
definition in \cite{lee}).  
Let us assume that class II operators up to and including the order
$1/m^{n-1}$ have already been removed from the Lagrangian ($n \geq 1$, at 
the beginning of the iteration, no class II operators have been removed
and $n=1$). Collect the class II operators of order $1/m^n$ in the form
\begin{equation}
        O_{\hbox{\it class II}} = \frac{1}{m^n} 
      \sibar (i\vD A^n + \overline{A^n} i\vD) \si.
\end{equation}
Since $P_+ \si = \si$ where $P_+$ is the projection ${1 + \slash v \over 2}$,
this operator may be rewritten
\begin{equation}
        O_{\hbox{\it class II}} = \frac{1}{m^n} 
        \sibar (i\vD P_+ A + \overline A P_+ i\vD) \si.
\end{equation}
Now perform the field redefinition
\begin{equation}
        \si \rightarrow (1 - \frac{1}{m^n} P_+ A^n) \si.
\end{equation}
The $P_+$ factor is there so that the identity $\slash v \si = \si$ may
be satisfied by the redefined fields as well as the original fields. It
is only necessary to include it if $A$ is not written to commute with
$P_+$ to begin with. 
Clearly, this field redefinition removes class II operators from the
Lagrangian up to and including $O(1/m^n)$. By iterating this process,
class II operators can be removed to any desired order.

Note that this field redefinition induces new terms at higher orders in
the Lagrangian. In general, we can split $A^n$ in the form
\begin{equation}
   A^n = A_1^n + A_2^n i D\cdot v 
\end{equation}
where $A_1^n$ does not vanish when operating on an on-shell field with
$i v D \si = 0$.
If $A_1^n$ is non-zero, the field redefinition to remove the class II
operators at order $1/m^n$ will change the coefficients of the class I
operators at order $1/m^{n+1}$, and therefore in general it is not
correct to simply throw out the class II operators from the Lagrangian.
As $A_1^1 = 0$, this did not happen in the present case of $1/m^2$, but
the naive method will yield incorrect results at order $1/m^3$.

\section{Eliminating class II operators during renormalization}

In general, even after the class II operators have been eliminated by
field redefinition, a renormalization of the theory will induce class II
terms. The theory with class I operators alone cannot be renormalized
without a revision of what is meant by a renormalization. Therefore,
when calculating without class II operators, an infinitesimal
renormalization must be followed by an infinitesimal field
redefinition that removes the class II operators. Then, in general, the
Lagrangian will possess the form
\begin{equation}
        {\cal L} = i\sibar\vD\si + C_{kin} O_{kin} + C_{mag} O_{mag} +
        C_1 O_1 + C_2 O_2,
\label{eq:lagrangian}
\end{equation}
to order $1 \over m^2$.

The renormalization of the order $1 \over m$ operators at one loop is
well known \cite{falk,kilian}.  $C_{kin}$ is not renormalized at
all, and
\begin{eqnarray}
        \beta_{mag} &=& \mu {\partial \over \partial \mu} C_{mag}
                = {6g^2 \over (4\pi)^2} C_{mag} \nonumber \\
                &=& {2Ng^2 \over (4\pi)^2} C_{mag}
                \quad \hbox{for $SU(N)$.}
\end{eqnarray}

The coefficient of $O_{\vD}$ is also renormalized at one loop.  However,
the renormalization is entirely multiplicative, so if this operator is
eliminated from the Lagrangian by a field redefinition after tree-level
matching, there will be no $O_{\vD}$ term after an infinitesimal
renormalization either.

C.L.Y. Lee \cite{lee} attempted to renormalize the order $1 \over m^2$
operators, but Lee's division of operators into class I and class II
parts was not Hermitian, so it was not actually possible to perform the
field redefinition necessary to eliminate the class II terms. Here we
renormalize the Lagrangian to order $1 \over m^2$ and one loop using the
Hermitian class I operator definitions (\ref{eq:operators}).  Also,
unlike Lee, we define coefficients only for the local operators in the
Lagrangian and derive $\beta$ functions in terms of those coefficients.
In \cite{Bal94}, S. Balk {\em et al.} discuss 
the renormalization of a variant of HQET constructed 
from a sequence of Foldy-Wouthuysen transformations to $O(1/m^2)$.
While the operator $O_1$ is not present in their basis, their results
seem to be consistent with ours regarding the running of $O_2$. 
For a comparision with other, more recent calculations 
\cite{balzereit,Blo96} see section~\ref{sec:bal}.

Since the class II terms induced by renormalization are all of order $1
\over m^2$, when renormalizing the Lagrangian, there is no need to
actually calculate the renormalization of all of the class II terms and
calculate the field redefinition necessary to remove them during
renormalization.  The effect of the field redefinition on the Lagrangian
to order $1 \over m^2$ will be equivalent to simply throwing out the
class II terms induced by renormalization.

The class I operators induce various Feynman vertices, shown in Figures
\ref{fig:kinmagverts} and \ref{fig:onetwoverts}.
The Feynman rules are given in the appendix. Under background field gauge,
as described in Abbott \cite{abbott}, explicit gauge invariance establishes
relations between the coefficients of the various vertex terms that are
preserved under renormalization; the vertices arising from a single gauge-
invariant operator are multiplied by the same running coupling constant.

\section{Renormalizing $O_2$ to one loop}

Because of this explicit gauge invariance, in order to renormalize an operator,
it is only necessary to examine the part of each 1PI loop diagram's
divergent part which has the form of the simplest of the operator's
Feynman vertices.  Furthermore, if the vertex factor in question has multiple
terms, it is only necessary to look for one of them, provided that it is
not possible to produce the sought-after term using class II operators.
This immensely simplifies the task of renormalizing the operators.

$O_2$ is the easier of the order $1 \over m^2$ operators to renormalize.
$O_2$'s one-gluon vertex has a term which is rather difficult to create
in diagrams involving the other operators:
\begin{equation}
        {g \over 4m^2}
        k'_{\alpha} p_{\beta} \sigma^{\alpha \beta} v^{\mu} T_a .
        \label{eq:o2vertex}
\end{equation}
This is a natural thing to look for, because it puts strict constraints
on the form of a divergence, so few diagrams will
contribute to its renormalization. For any divergence to renormalize
this interaction, it must not only have the same tensor structure,
but also depend on the external momenta $k'$ and $p$ in the correct
way. They must both be contracted with $\sigma^{\alpha \beta}$, rather
than with $v$, with themselves, or with each other.

\subsection{Diagrams with one internal gluon line}
It is worthwhile to consider a very common situation in which these
criteria are usually {\it not} satisfied. Suppose that a one-loop
diagram has only one internal gluon line. It is perfectly legitimate to
label the momenta so that the loop integration variable, $q$, is the
momentum of that internal gluon. Then, no factors of $p$ and
no factors of $k'$ will appear in the gluon propagator.

There will, in general, be terms with factors of $k'$ and $p$ in the
denominators of the heavy quark propagators. However, there a momentum
always appears contracted with the heavy quark's four-velocity $v$.
Therefore, expanding these propagators in powers of $p$ will never yield
a term in which $k'$ or $p$ is contracted with anything other than $v$.

When doing a loop integral, the typical procedure is to combine
denominators using some sort of Feynman parameter, then shift the
integration variable so as to make the Euclidean integral hyperspherically
symmetric. In this case, though, all the propagators but one are heavy
quark propagators. The shift in the integration variable
only transforms $q$ into $q-\lambda v$, where $\lambda$ is a Feynman
parameter with dimensions of mass. Factors of $q$ do not turn into
factors of $k'$ or $p$ later in the calculation.

This would seem to indicate that in a diagram with only one internal
gluon line, any factor of (\ref{eq:o2vertex}) in
the divergence must explicitly appear in the product of the vertex
factors before the loop integral is done. However, there is another
possible complication
\cite{private}. Integrating over all momentum space can
transform products of loop momenta into factors of the metric. In
particular, if $f(q^2)$ is any scalar function that depends on $q^\mu$
only via $q^2$,
\begin{equation}
	\qint q^\mu q^\nu f(q^2) = \qint {1 \over 4} g^{\mu \nu} q^2 f(q^2)
	\label{eq:qqmetric}
\end{equation}
When considering which diagrams have the correct contractions to
contribute to the renormalization of $O_2$, it is necessary to take into
account situations in which things that ought to be contracted with each
other are both contracted with the loop momentum.

Fortunately, the terms we are looking for already have two factors of
the external momenta in them, so this mechanism requires the presence of
at least four factors of momentum in the numerator. Therefore, it only
operates in a few diagrams. (In fact, it turns out not to lead to any
new diagrams with one internal gluon line in the renormalization of
$O_2$, because there is no way to get four factors of momentum from the
requisite sets of vertices.)

These considerations allow the elimination of many diagrams with little
effort.

\subsection{Diagrams with two internal gluon lines}
The vast majority of relevant diagrams therefore contain two internal
gluon lines, and a three-gluon QCD vertex connected to the external
gluon leg. In these diagrams, factors
of $p$ may arise from places other than the various vertices involving
quarks. $p$ appears in the three-gluon QCD vertex. It also appears in
at least one of the internal gluon momenta, which makes it easier to
obtain it in the quark-gluon vertex factors.

It is still not necessary to worry about factors of $p$ in the propagator
denominators when renormalizing $O_2$, since at worst they will produce
factors of $p \cdot v$ and $p^2$, not factors of $p$ contracted with
the $\sigma$ tensor.

Any factors of $k'$ must {\it still} appear explicitly in the quark-gluon
vertex factors prior to loop integration. It is possible to label the
momenta in the loop so that $k'$ only goes through heavy quark lines.
In denominators it is contracted only with $v$, and it never appears in
the expression for the shift in $q$ after combining denominators.

However, now the identity (\ref{eq:qqmetric}) is important, because
it is possible to obtain four factors of momentum with the use of the three-
gluon QCD vertex in addition to the heavy quark vertices. It is necessary
to consider both diagrams in which (\ref{eq:o2vertex}) appears
explicitly in the numerator prior to loop integration, and diagrams in
which some of the factors in (\ref{eq:o2vertex}) are contracted with
$q^\mu q^\nu$ rather than with each other.

These facts allow the elimination of some of the diagrams with two
internal gluons. Diagrams with more internal gluons do not appear,
because they have either too many loops or too many legs.

\subsection{Operator insertions that do not contribute}
The term (\ref{eq:o2vertex}) is spin-dependent, so it is not necessary
to consider diagrams containing only spin-independent vertices. Double
insertions of $O_{kin}$, and insertions of $O_1$, will therefore not contribute
to the running of $O_2$.
Either $O_{mag}$ or $O_2$ itself has to be in the diagram somewhere.

Double insertions of $O_{mag}$ cannot produce (\ref{eq:o2vertex}) either,
because $O_{mag}$ has no factors of $k'$ in any of its vertices.
Therefore, the only contributions to the renormalization of $O_2$
must come from diagrams involving an insertion of $O_{kin}$
{\it and} $O_{mag}$, or diagrams with a single insertion of $O_2$.

In this case, there is also no need to subtract out divergences of the
form of the class II operators, because none of the class II operators
has a one-gluon vertex with a term of the same form as
(\ref{eq:o2vertex}). The only spin-dependent class II operator of order
$1 \over m^2$ is ${1\over 8m^2} \sibar \lbrace iD \cdot v, G^{\alpha
\beta} \sigma_{\alpha \beta} \rbrace \si$, and the heavy quark residual
momentum $k'$ is never contracted with $\sigma$ in the vertices.

\subsection{Diagrams with $O_{kin}$ and $O_{mag}$}

There are many one-loop diagrams involving one $O_{kin}$ vertex and one
$O_{mag}$ vertex; fortunately, most of them do not contribute to the
term in question.

As said above, in all cases the correct factor of $k'$ must arise
explicitly from
the vertex factors, and they must be contracted as in (\ref{eq:o2vertex}),
except that factors of $q^\mu q^\nu$ may appear in place of the metric.

If the one-gluon vertex of $O_{kin}$ appears, that vertex may
not be connected to the external gluon leg or to a leading-order quark-
gluon vertex, because in neither case will $k'$ end up contracted with
the $\sigma$ tensor in $O_{mag}$, or with the loop momentum $q$. It must
be connected to an internal gluon line.

If the gluonless $O_{kin}$ vertex appears on an internal quark line
carrying some linear combination of $k'$, $p$, and $q$, the only way
(\ref{eq:o2vertex}) may be obtained is in the term with a factor of $k'
\cdot q$, so that (\ref{eq:qqmetric}) can leave $k'$ contracted with
$\sigma$.

In either case, the $k'$ term in the $O_{kin}$ vertex has no factor of $p$.
Therefore, a factor of $p$ must appear somewhere else.
If there is only one internal gluon line, then the factor of $p$ must
come from the one-gluon $O_{mag}$ vertex, by the general argument above. But
it cannot do so. If the momentum in the $O_{mag}$ vertex is $p$, then
that vertex must be connected to the external leg, and there is no
way to contract the factor of $\sigma$ with $k'$ or $q$.

\begin{figure}
\begin{center}
\mbox{\epsfig{figure=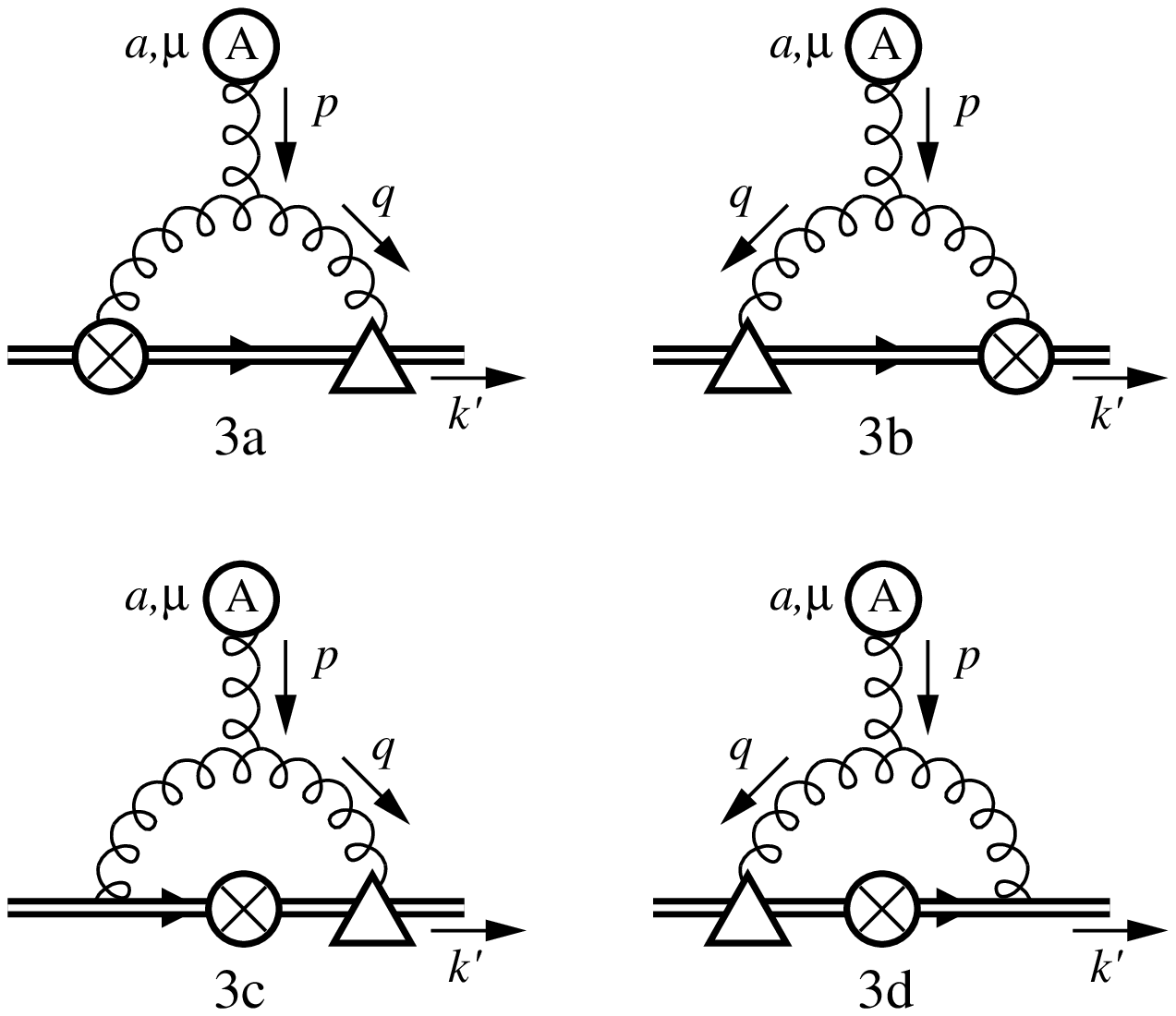,height=8cm}}
\end{center}
\caption{Diagrams involving $O_{kin}$ and $O_{mag}$
that renormalize $O_2$.}
\label{fig:kinmagtwo}
\end{figure}

The diagram must, therefore, have two internal gluon lines. If $k'$ or
$q$ is to end up contracted with the $\sigma$ tensor, an internal gluon
line has to contract with $O_{mag}$. There are only four such diagrams,
shown in Figure \ref{fig:kinmagtwo}. The calculation of the contribution
of Figures \ref{fig:kinmagtwo}a-b is described in more detail than the
other parts of the calculation; it is a typical example of the process of
calculating these divergences.

The amplitudes corresponding to these diagrams are
\begin{eqnarray}
        &\int {d^4 q \over (2\pi)^4}& g f_{abc}
        [g_{\mu\kappa}(-2p_\rho)+g_{\rho\kappa}(2q-p)_\mu+
        g_{\mu\rho}(2p_\kappa)] \nonumber \\
        & \times & {-i \over (p-q)^2 q^2 (k'-q)\cdot v} \nonumber \\
        & \times & \left(-g \over 2m \right)
        \sigma^{\kappa\alpha} q_\alpha T_c
        \left( ig \over 2m \right) (2k'-q-p)^\rho T_b
\end{eqnarray}
from the diagram in Figure \ref{fig:kinmagtwo}a, and
\begin{eqnarray}
        &\int {d^4 q \over (2\pi)^4}& g f_{abc}
        [g_{\mu\kappa}(-2p_\rho)-g_{\rho\kappa}(2q-p)_\mu+
        g_{\mu\rho}(2p_\kappa)] \nonumber \\
        & \times & {-i \over (p-q)^2 q^2 (k'-p+q)\cdot v} \nonumber \\
        & \times & \left(-g \over 2m \right)
        \sigma^{\rho\alpha} q_\alpha T_c
        \left( ig \over 2m \right) (2k'-q-p)^\kappa T_b
\end{eqnarray}
from Figure \ref{fig:kinmagtwo}b. The projection operator in the
heavy quark propagator may be ignored, since all operators are
assumed to be sandwiched between heavy quark fields, and in both
cases the projector commutes with one of the vertex factors.

Taking only the terms with a single factor of $k'$
contracted with $\sigma$, and adding the two diagrams together, gives
\begin{equation}
        -2 Ni {g^3 \over (2m)^2} T_a k'_\beta \sigma^{\beta\alpha}
        \int {d^4q \over (2\pi)^4} {q_\alpha (2q-p)_\mu \over (p-q)^2 q^2}
        \left( {1 \over q \cdot v} + {1 \over (q-p)\cdot v } \right)
\end{equation}
where $N$ is the number of colors. The quadratic factors in the
denominator may be combined using the usual method of Feynman
parameters; shifting the integral to make it symmetric, and extracting
only the term linear in $p$ yields
\begin{equation}
        -2Ni {g^3 \over (2m)^2} T_a k'_\beta p_\alpha \sigma^{\beta \alpha}
        \int {d^4q \over (2\pi)^4} {q_\mu\over (q^2)^2 q\cdot v}
\end{equation}
The heavy quark propagator factors remaining in the denominator
may now be combined with the rest using the usual (for HQEFT) identity
\begin{equation}
        {1 \over (q^2)^n(q\cdot v)^m}
        = {(n+m-1)! \over (n-1)! (m-1)!}
        \int^\infty_0 {2^m \lambda^{m-1} d \lambda
        \over (q^2 + 2 \lambda q \cdot v)^{n+m} }
\end{equation}
where $\lambda$ is a sort of Feynman parameter with dimensions of mass.
Then the integral over momentum space is
\begin{equation}
        \int^\infty_0 d\lambda \int {d^4q \over (2\pi)^4}
        4 {q_\mu \over (q^2+2 \lambda q \cdot v)^3}
\end{equation}
Again shift the integral, $q \rightarrow q-\lambda v$,
\begin{equation}
        \int^\infty_0 d\lambda \int {d^4q \over (2\pi)^4}
        4 {(q-\lambda v)_\mu \over (q^2-\lambda^2)^3}
\end{equation}
and remove the term that is odd in $q$; now the integral over $\lambda$
is easy, and the amplitude becomes
\begin{equation}
        -2Ni {g^3 \over (2m)^2} T_a k'_\beta p_\alpha \sigma^{\beta \alpha}
        v_\mu \int {d^4q \over (2\pi)^4} {1 \over (q^2)^2}
\end{equation}
Under $DR\overline{MS}$ in $4-\epsilon$ dimensions,
the amplitude is
\begin{equation}
        {4N g^2 \mu^\epsilon \over (4\pi)^2 \epsilon}
        \cdot {g \over 4m^2}
        k'_{\alpha} p_{\beta} \sigma^{\alpha \beta} v^{\mu} T_a.
\end{equation}

In Figures \ref{fig:kinmagtwo}c-d, contributions of the form of
(\ref{eq:o2vertex}) arise from factors of the form $k' \cdot q
q_\alpha p_\beta \sigma^{\alpha \beta}$, via (\ref{eq:qqmetric}).
Taken together, the relevant part of their contribution is
\begin{equation}
	{g^3 \over 4m^2} 4NiT_a v^\mu \sigma^{\gamma \alpha}
	p_\gamma k'^\beta \qint {q_\alpha q_\beta \over (q^2)^2 (q \cdot v)^2}
\end{equation}
After using Feynman parameters to combine the denominators and making
the integrand symmetric, (\ref{eq:qqmetric}) may be used to obtain a
term of the desired form, which is
\begin{equation}
        {-8N g^2 \mu^\epsilon \over (4\pi)^2 \epsilon}
        \cdot {g \over 4m^2}
        k'_{\alpha} p_{\beta} \sigma^{\alpha \beta} v^{\mu} T_a.
\end{equation}

\subsection{Diagrams with $O_2$}

Some of the diagrams containing an insertion of $O_2$ may be eliminated
in an analogous manner to the ones considered above. Here it is not
necessary to take (\ref{eq:qqmetric}) into consideration, because an
$O_2$ vertex and a three-gluon QCD vertex can together contribute at
most three factors of momentum in the numerator.

The three-gluon $O_2$ vertex is completely nonderivative,
so there is no way to obtain the desired factor of $k'$ from that vertex.

\begin{figure}
\begin{center}
\mbox{\epsfig{figure=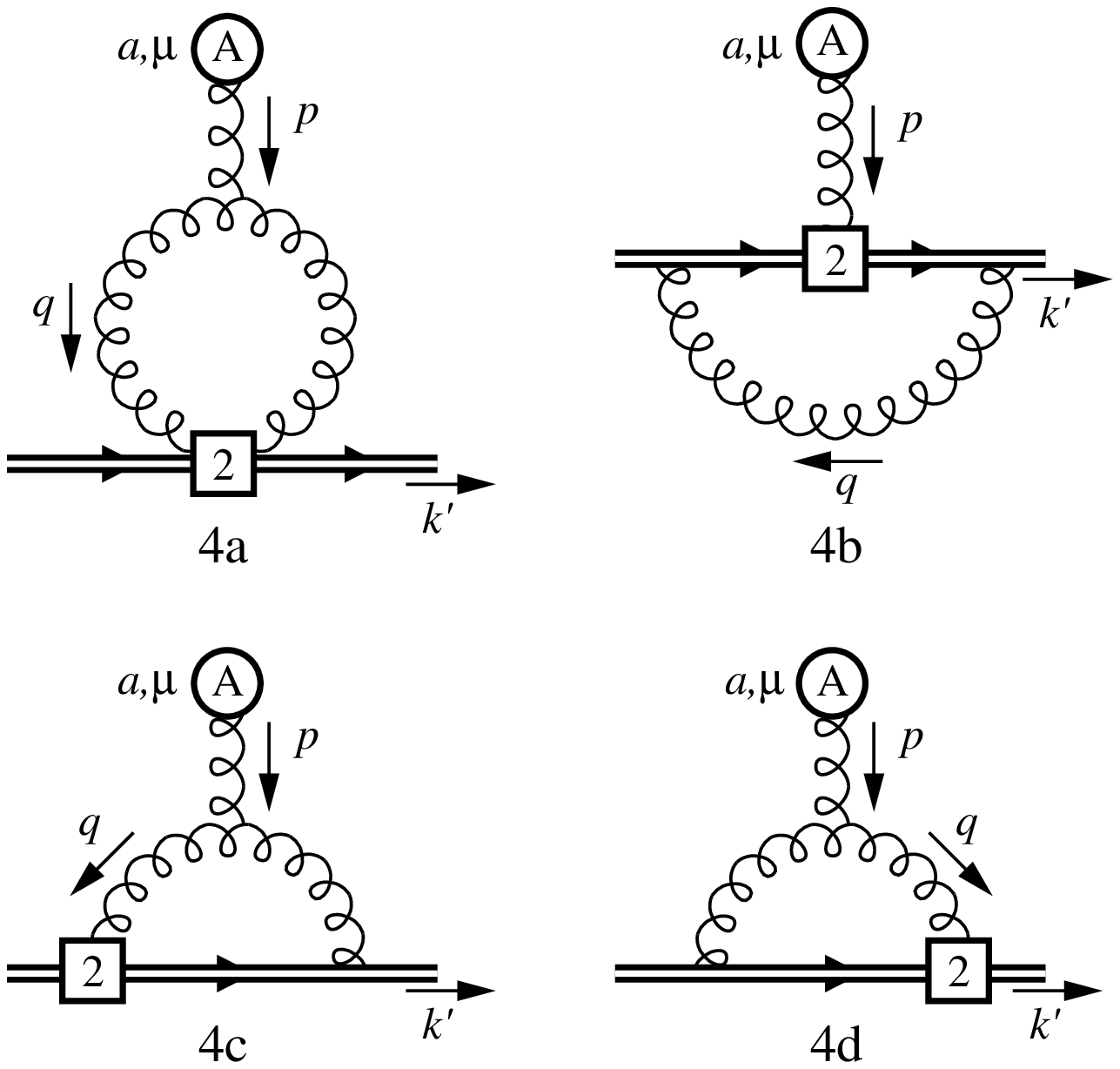,height=8cm}}
\end{center}
\caption{Diagrams involving $O_2$ that renormalize $O_2$.}
\label{fig:twotwo}
\end{figure}

The term in the two-gluon vertex that has the $k'$ lacks any other
factors of momentum, so the factor of $p$ must come from somewhere else.
Therefore, any such diagram must have two internal gluon lines. The only
possibility is Figure \ref{fig:twotwo}a.

With the one-gluon vertex, there will be no contribution from diagrams
in which $O_2$ is connected to the sole internal gluon line, since then
the gluon momentum in the vertex is $q$, and there are no other vertices
with momenta in them.  However, there is also
the option of connecting $O_2$ directly to the external gluon leg, so
that a factor of the desired form comes directly from the $O_2$ vertex.
That is Figure \ref{fig:twotwo}b.

There are also diagrams with two internal gluon lines, one of which
contracts with $O_2$'s one-gluon vertex; these are Figure \ref{fig:twotwo}c
and Figure \ref{fig:twotwo}d.

The calculations proceed much as before.  In each diagram, the terms to
look for are the ones in which the momenta are appropriately contracted,
keeping in mind that factors of a loop momentum $q$ can turn into
factors of $p$ when the integration variable is shifted, if $p$ flows
through an internal gluon line (but {\it not} if it only flows through
an internal heavy quark line).  The important term contributed by
Figure \ref{fig:twotwo}a is
\begin{equation}
        {1 \over 2} \cdot
\qint {g^3 \over (2m)^2} NiT_a {1 \over q^2 (p-q)^2}
        (4k_\alpha p_\rho \sigma^{\alpha \rho} v_\mu)
\end{equation}
The $1 \over 2$ is the loop integral's symmetry factor. Combining
denominators in the usual way reveals that the $p$ in the denominator
contributes only terms of higher than linear order in $p$ and may
therefore be neglected (the story would have been different, had there
been terms with factors of $q$ in the numerator). The divergent term
linear in $p$ is
\begin{equation}
        -{4N g^2 \mu^\epsilon \over (4\pi)^2 \epsilon}
        \cdot {g \over 4m^2}
        k'_{\alpha} p_{\beta} \sigma^{\alpha \beta} v^{\mu} T_a.
\end{equation}
Figure \ref{fig:twotwo}b makes a small contribution, because of a group-theoretic
factor of $-1 \over 2N$; the term with the correct factors of $k$ and
$p$ is
\begin{equation}
        {-ig^3 \over (2m)^2} \left(-{1 \over 2N} T_a\right)
        k'_\alpha p_\beta \sigma^{\alpha \beta} v^\mu
        \qint {1 \over q^2(q\cdot v)^2},
\end{equation}
which comes to
\begin{equation}
        +{2 g^2 \mu^\epsilon \over N (4\pi)^2 \epsilon}
        \cdot {g \over 4m^2}
        k'_{\alpha} p_{\beta} \sigma^{\alpha \beta} v^{\mu} T_a.
\end{equation}
Figures \ref{fig:twotwo}c and \ref{fig:twotwo}d contribute in much the manner of
Figures \ref{fig:kinmagtwo}a-b.  The term of the desired form in Figure \ref{fig:twotwo}c,
after combining the gluon denominators, shifting the loop variable, and
doing the integral over the Feynman parameter, is
\begin{equation}
        -{Ni\over 2} T_a {g^3\over 2m^2}
        k'_\alpha p_\beta \sigma^{\alpha \beta}
        \qint{2 q\cdot v v_\mu - q_\mu \over (q^2)^2 q\cdot v}
\end{equation}
Figure \ref{fig:twotwo}d's contribution is identical.  Adding the two together,
combining with the heavy quark denominators and performing the requisite
integrals gives
\begin{equation}
        +{2N g^2 \mu^\epsilon \over (4\pi)^2 \epsilon}
        \cdot {g \over 4m^2}
        k'_{\alpha} p_{\beta} \sigma^{\alpha \beta} v^{\mu} T_a.
\end{equation}

\subsection{The $\beta$ function of $O_2$'s coefficient}

These, then, are all of the divergent pieces of $\Gamma_{O_2}$,
the part of the 1PI three-point function with the same form as (18).
Adding them all together, including the coefficients from the
Lagrangian, and putting in the contribution from the single vertex,
\begin{eqnarray}
        \Gamma_{O_2} &=&
        \left\lbrace C_2 + \left[-4NC_{kin}C_{mag}
        -\left(2N - {2 \over N}\right) C_2 \right]
        {g^2 \over (4\pi)^2} \ln \mu \right\rbrace \nonumber \\
        &\times& {g \over 4m^2}
        k'_{\alpha} p_{\beta} \sigma^{\alpha \beta} v^{\mu} T_a + \cdots
\end{eqnarray}
where the dots represent convergent terms not dependent on $\mu$.
Now the $\beta$ function for $C_2$ may be determined by solving the
renormalization group equation for $\Gamma_{O_2}$:
\begin{equation}
        \left(  \mu {\partial \over \partial \mu}
             +  \beta_g {\partial \over \partial g}
             +  \beta_{mag} {\partial \over \partial C_{mag}}
             +  \beta_{kin} {\partial \over \partial C_{kin}}
             +  \beta_2 {\partial \over \partial C_2}
             -  2 \gamma_Q - \gamma_A \right) \Gamma_{O_2} = 0
\end{equation}
$\beta_{kin}$ is zero ($C_{kin}$ does not run and is just equal to 1),
the $\beta_{mag}$ term is higher order in $g$ than the others, and
explicit gauge invariance in background field gauge means that
the terms with $\beta_g$ and $\gamma_A$ cancel.  The anomalous
dimension of the heavy quark field is
\begin{equation}
        \gamma_Q =-\left( N - {1 \over N} \right) {g^2 \over (4\pi)^2}.
\end{equation}
Solving for $\beta_2$ to order $g^2$,
\begin{equation}
        \beta_2 (g(\mu), C_{kin}, C_{mag}(\mu))
        = {g^2 \over (4 \pi)^2}
        \left( 4NC_{kin}C_{mag} \right)
\end{equation}
Remarkably, there is no multiplicative renormalization of $C_2$; the
diagrams with $O_2$ insertions are completely canceled by the term from
the heavy quark anomalous dimension.
At the scale where matching occurs, $C_{mag} = C_{kin} = C_2 =1$, and
\begin{equation}
        \beta_2 (g(m_Q), C_{kin}, C_{mag}(m_Q))
        = {4N g^2 \over (4 \pi)^2}
\end{equation}

Since every diagram except \ref{fig:twotwo}b contains a three-gluon QCD
coupling, it is also interesting to consider the case of a $U(1)$ gauge
theory, in which \ref{fig:twotwo}b is the only 1PI diagram that
renormalizes $O_2$. Then \ref{fig:twotwo}b lacks the group-theoretic
factor of $-1\over 2N$ that is present in the $SU(N)$ case, and the
heavy quark anomalous dimension is $2g^2 \over (4\pi)^2$. The solution
of the RGE to order $g^2$ is
\begin{equation}
        \beta_2 = {g^2 \over (4 \pi)^2} (+4-4) C_2 = 0
\end{equation}
so, for $U(1)$, $C_2$ does not run at all at one loop.

\section{Renormalizing $O_1$ to one loop}

\subsection{Eliminating class II terms}
The renormalization of $O_1$ is slightly more involved, conceptually and
mathematically, for two reasons.  First, there are more diagrams to
consider; second, the simplest vertex of $O_1$ has no term that cannot
be produced by class II operators as well, so it is necessary to
calculate the part of each diagram that renormalizes one of the
class II operators, and subtract it out.  Fortunately, these two
difficulties cancel each other out to some extent, because many of the
extra diagrams turn out to renormalize only the class II operator.

The term in the one-gluon vertex of $O_1$ that gives the least trouble is
\begin{equation}
        {ig \over 8m^2} (p^2 v^{\mu}) T_a.
        \label{eq:o1vertex}
\end{equation}
This term also appears in the one-gluon vertex of the class II operator
${1 \over 8m^2} \sibar \lbrace iD \cdot v, D^2 \rbrace \si$. However, in
this operator the factor $p^2 v^\mu$ always shows up in the combination
$(-2k' \cdot p + p^2) v^\mu$, and the $k' \cdot p v^\mu$ term is not produced
by any other local operator to order $1 \over m^2$. Therefore, to
subtract out the renormalization of the class II operator, all that is
necessary is to add, in the divergent term of each diagram, {\it half}
the coefficient of
\begin{equation}
        {ig \over 8m^2} (k' \cdot p v^{\mu}) T_a
        \label{eq:classIIvertex}
\end{equation}
to the coefficient of (\ref{eq:o1vertex}).

Both of these terms are spin-independent.  Therefore, contributions to
the renormalization may come from double insertions of $O_{kin}$, from
double insertions of $O_{mag}$, or from single insertions of $O_1$
itself.

\subsection{Diagrams with one internal gluon line}

The expression (\ref{eq:o1vertex}) has no factor of $k'$ in it, so it is
now possible to obtain nonzero contributions to the running of $O_1$
from diagrams in which no factors of $k'$ appear in the vertices.
There may be a contribution if the numerator, prior to loop integration,
contains factors like $k' \cdot p$, $p^2$, $k' \cdot q p \cdot q$, or
$p \cdot q p\cdot q$. If there is only one internal gluon line, these
factors cannot arise from the propagators, for precisely the same reasons
as in the $O_2$ case. Routing the loop momentum $q$ through the lone gluon
line reveals that its denominator is simply $q^2$, and factors of $k'$,
$p$, or $q$ in the quark denominators are contracted with $v$.

Therefore, in such diagrams, the factors of $k'$ and $p$ in
the divergence must appear explicitly in the vertex factors prior to
loop integration, and the external momenta must be contracted with
each other or with $q^\mu q^\nu$.

Class II contributions may also be neglected. If the factors of the
form of $p^2$ and $k' \cdot p$ appear in
the combination $p^2 - 2k' \cdot p$, then the divergence just
renormalizes a class II operator, and it need not be considered.

\subsection{Diagrams with two internal gluon lines}

When there are two internal gluon lines, additional complications arise.
Factors of $p^2$ can and do arise in the gluon propagators. Therefore,
when calculating the divergent term of the form of (\ref{eq:o1vertex}),
it is necessary to expand the loop integrand in powers of $p^2$ and
retain the zeroth- and first-order terms. For this reason, the
calculation of these diagrams is more involved than for $O_1$.

\subsection{Diagrams with $O_{kin}$}

Terms like (\ref{eq:o1vertex}) and (\ref{eq:classIIvertex}) arise from
many diagrams with two $O_{kin}$ vertices.

\subsubsection{The gluonless vertex}

In particular, it seems as
if contributions to both ought to come from any diagram with a no-gluon
vertex on a quark line carrying some linear combination of $k'$ and $p$
(and possibly the loop momentum $q$). This is indeed the case for the
diagrams with two internal gluon lines (as in Figures \ref{fig:kinkinone}b-d).

If there is only one internal gluon line, the possibilities are more
restricted. If there is also only one gluonless vertex, then there can be at
most three factors of momentum and (\ref{eq:qqmetric}) cannot provide
the desired contractions. Routing $q$ through the gluon line implies
that the quark line on which the $O_{kin}$ vertex sits may
only carry either $k'\pm q$ or $k' - p\pm q$. In the former case, the
gluonless $O_{kin}$ vertex provides neither a term like
(\ref{eq:o1vertex}) nor one like (\ref{eq:classIIvertex}). In the latter
case, the terms appear in the class II combination and may be ignored.

If there is one internal gluon line but two gluonless $O_{kin}$ vertices,
it becomes possible to obtain contributions via (\ref{eq:qqmetric}).
If both of them are on the same side of the external gluon leg, however,
the momentum factors are of the same form as those of a single vertex, and
the diagrams may be excluded by the arguments in the previous paragraph.
The only class I contribution occurs when the gluonless vertices are
on opposite sides of the external gluon leg, as in Figure \ref{fig:kinkinone}f.

\subsubsection{The one-gluon vertex}

The one-gluon vertex has uncontracted momenta in it; factors of $k'
\cdot p$ and $p^2$ may be produced by putting it in a diagram with two
internal gluon lines (Figures \ref{fig:kinkinone}a-c) or by contracting
it with another one-gluon $O_{kin}$ vertex (Figure \ref{fig:kinkinone}e).

\subsubsection{Eliminating the two-gluon vertex}

The two-gluon vertex is completely nonderivative, so any diagram with
that one in it would have to get its factors of $k' \cdot p$ or $p^2$
from somewhere else. If there is only one internal gluon line, then,
as stated above, these factors would have to come from the other $O_{kin}$
vertex. They can't, since they could only arise from the no-gluon
vertex, and that only supplies them in a class II combination if there
is only one internal gluon line. Therefore, the two-gluon $O_{kin}$
vertex can only contribute if there are two internal gluon lines.

But if the diagram is to be 1PI, then one of those has to connect
to something other than the two-gluon $O_{kin}$ vertex, which
leaves it with an extra leg. So this vertex doesn't contribute at
all.

\begin{figure}
\begin{center}
\mbox{\epsfig{figure=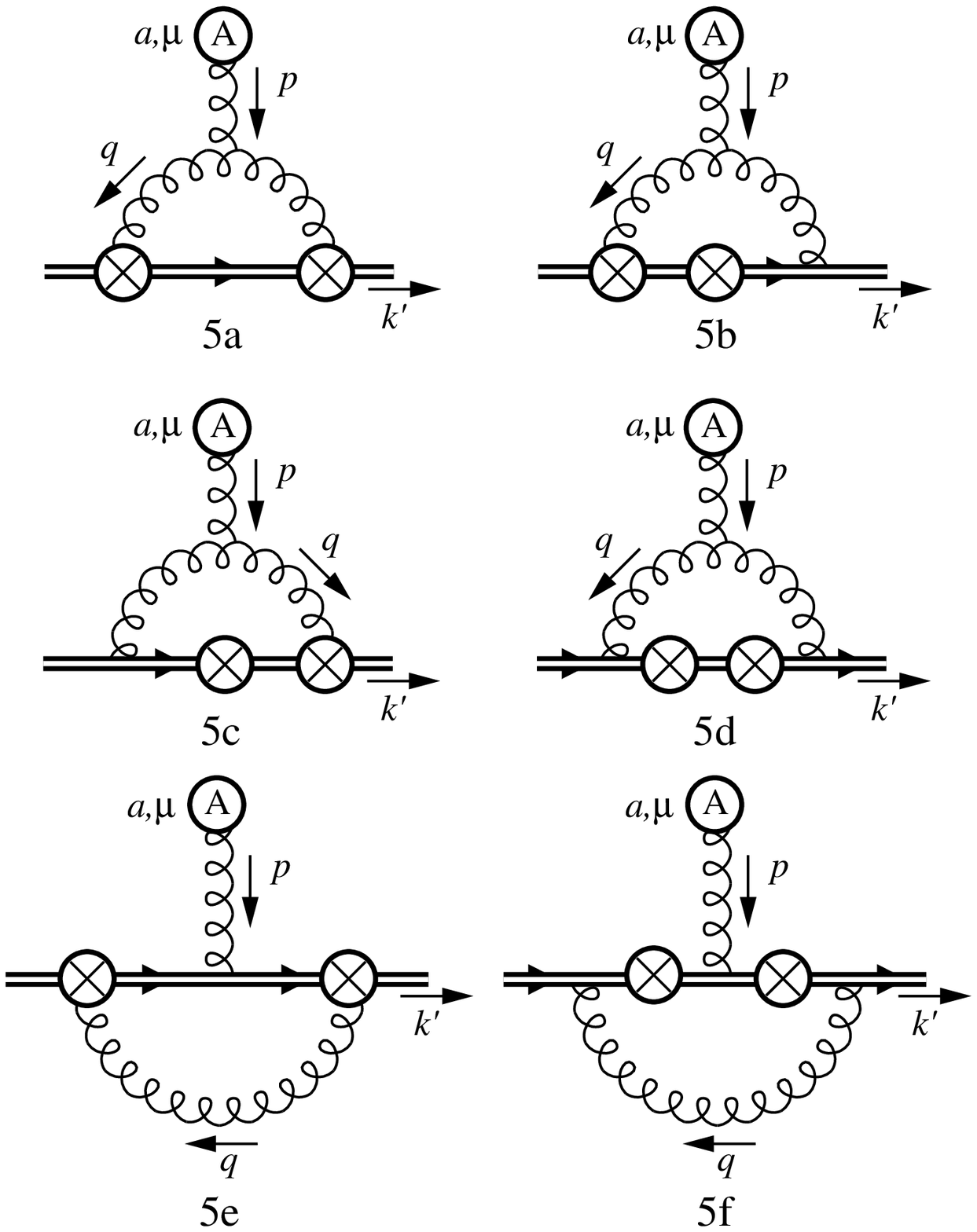,height=15cm}}
\end{center}
\caption{Diagrams involving $O_{kin}$ that renormalize $O_1$.}
\label{fig:kinkinone}
\end{figure}

The diagrams in Figure \ref{fig:kinkinone} are therefore all of the
contributions to the renormalization of $O_1$ from double insertions of
$O_{kin}$. When calculating them, it is necessary to do the expansions
in $p^2$ mentioned above.

For example, consider Figure \ref{fig:kinkinone}a.  After the usual
manipulations to combine the gluon denominators, shift the integrand,
and throw out terms proportional to $k'^2$ or $k' \cdot v$ or to
four-vectors other than $v^\mu$ or $q^\mu$, the leftover amplitude looks
like
\begin{eqnarray}
        & - & {g^3 \over (2m)^2} {N T_a \over 2} \int^1_0 dx \qint
        {1 \over [q^2 + x(1-x)p^2]^2 (q \cdot v)} 2 q^\mu \nonumber \\
        & \times & [(x^2+x+1) p^2 + q^2 + (1+2x) p\cdot q
        - 4k'\cdot q -2(1+2x)k'\cdot p]
\end{eqnarray}
Expanding the factor with the denominators
\begin{equation}
        {1 \over [q^2 + x(1-x)p^2]^2} \rightarrow
        {1 \over (q^2)^2} - {2 x(1-x) p^2 \over (q^2)^3} + \cdots
\end{equation}
reveals the presence of additional $p^2$ terms in the result.  The
relevant divergent terms turn out to be
\begin{equation}
        {g^3 \over (2m)^2} T_a \left(4N k' \cdot p +{N\over 2} p^2
        \right) v^\mu
        \qint {1 \over (q^2)^2}
\end{equation}
Subtracting out the class II contribution, by adding the coefficient
of the $p^2$ term to half the coefficient of the $k' \cdot p$ term, makes
this
\begin{equation}
        {10N g^2 \mu^\epsilon \over (4\pi)^2 \epsilon}
        {ig \over 8m^2} T_a p^2 v^\mu + (II)
\end{equation}
where (II) refers to the part that renormalizes an operator that
vanishes by the equation of motion.

Figures \ref{fig:kinkinone}b and \ref{fig:kinkinone}c together contribute
\begin{equation}
        -2N {g^3 \over (2m)^2} T_a (k' \cdot p - p^2) v^\mu
        \qint {1 \over (q^2)^2}
\end{equation}
which is
\begin{equation}
        {4N g^2 \mu^\epsilon \over (4\pi)^2 \epsilon}
        {ig \over 8m^2} T_a p^2 v^\mu + (II),
\end{equation}
and Figure \ref{fig:kinkinone}d contributes
\begin{equation}
        {g^3 \over (2m)^2} T_a (-8Nk' \cdot p +2N p^2) v^\mu
        \qint {1 \over (q^2)^2}
\end{equation}
which is
\begin{equation}
        {-8N g^2 \mu^\epsilon \over (4\pi)^2 \epsilon}
        {ig \over 8m^2} T_a p^2 v^\mu + (II).
\end{equation}
Figures \ref{fig:kinkinone}e-f have a different group-theoretic structure;
\ref{fig:kinkinone}e's contribution is
\begin{equation}
        -{16 \over N}{g^2 \mu^\epsilon \over (4\pi)^2 \epsilon}
        {ig \over 8m^2} T_a k \cdot p v^\mu
\end{equation}
which is
\begin{equation}
        -{8 \over N}{g^2 \mu^\epsilon \over (4\pi)^2 \epsilon}
        {ig \over 8m^2} T_a p^2 v^\mu + (II).
\end{equation}
and {fig:kinkinone}f's is
\begin{equation}
		\left(-{1 \over 2N}T_a \right) {g^3\over 4m^2} v^\mu
		\qint {q^2(p^2-2k' \cdot p) - 4 k' \cdot q p \cdot q \over
		q^2 (q \cdot v)^4}
\end{equation}
which, after combining denominators, shifting the integration variable, and
applying (\ref{eq:qqmetric}), becomes
\begin{equation}
        {8 \over 3N}{g^2 \mu^\epsilon \over (4\pi)^2 \epsilon}
        {ig \over 8m^2} T_a p^2 v^\mu + (II).		
\end{equation}
In this case the entire class I contribution comes from (\ref{eq:qqmetric}).

\subsection{Diagrams with $O_{mag}$}

Double insertions of $O_{mag}$ will produce no factors of $k' \cdot p$,
since in such diagrams the only factors of $k'$ are in heavy quark
propagators.  Therefore, it is only necessary to seek out factors of
$p^2 v^\mu$.

If the diagram has only one internal gluon line, at least one of the $O_{mag}$
vertices will be connected to it, so its vertex will contain a factor of
$q$ rather than $p$. Then there is no way of obtaining two factors
of $p$ explicitly in the quark-gluon vertex factors; (\ref{eq:qqmetric})
does not apply, since there is no way to obtain four factors of momentum
in the numerator.

\begin{figure}
\begin{center}
\mbox{\epsfig{figure=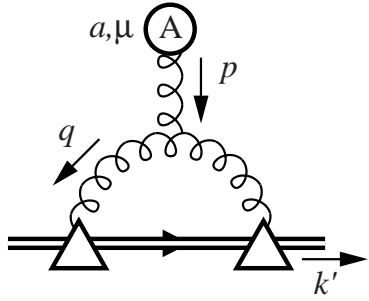,height=4cm}}
\end{center}
\caption{Diagram involving $O_{mag}$ that renormalizes $O_1$.}
\label{fig:magmagone}
\end{figure}

There must be two internal gluon lines. That leaves only Figure
\ref{fig:magmagone}.

In this calculation, the projector in the heavy quark propagator must not
be neglected, since it does not commute with $\sigma^{\mu \nu}$ \cite{private}.
When extracting the spin-independent part of this diagram's amplitude, one
may use the identities
\begin{equation}
	\sigma^{\alpha \mu} \left({1+\slash v \over 2}\right) \sigma_{\alpha \nu}
	= 2 \delta^\mu_\nu - 2 v^\mu v_\nu
	+ \hbox{spin-dependent terms}
\end{equation}
and
\begin{eqnarray}
	\sigma_{\mu \alpha} \left({1 + \slash v\over 2}\right)
	\sigma_{\beta \nu} p^\alpha p^\beta
	&=&  -g_{\mu \nu} p^2 + g_{\mu \nu}(p \cdot v)^2 + p_\mu p_\nu
	+ p^2 v_\mu v_\nu \nl
	&&- v_\mu p_\nu p \cdot v - p_\mu v_\nu p \cdot v
	+ \hbox{spin-dependent terms}
\end{eqnarray}
where both expressions are assumed to be sandwiched between heavy-quark
spinors, so that factors of $\slash v$ may be dropped at the beginning
or end of a term.

Figure \ref{fig:magmagone}'s contribution is
\begin{equation}
        {10N g^2 \mu^\epsilon \over 3(4\pi)^2 \epsilon}
        {ig \over 8m^2} T_a p^2 v^\mu.
\end{equation}
There is no class II term.

\subsection{Diagrams with $O_1$}
In the multiplicative renormalization from insertions of $O_1$, no $k'
\cdot p v^\mu$ terms arise, so there are no class II renormalizations to
subtract away. Again, (\ref{eq:qqmetric}) does not apply since there are
at most three factors of momentum in the numerator.

\begin{figure}
\begin{center}
\mbox{\epsfig{figure=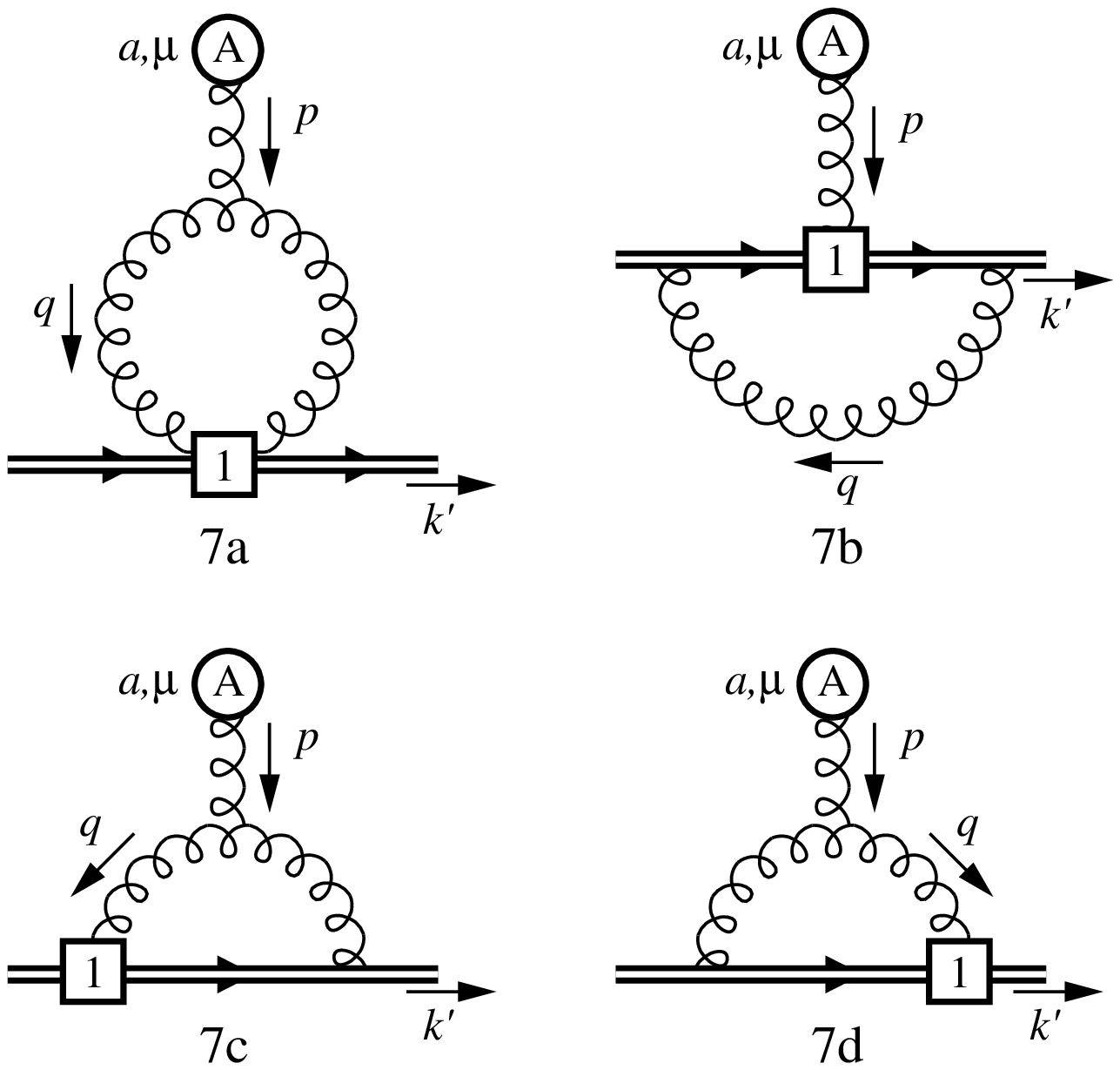,height=9cm}}
\end{center}
\caption{Diagrams involving $O_1$ that renormalize $O_1$.}
\label{fig:oneone}
\end{figure}

If there is only one internal gluon line and it contracts with any of
the $O_1$ vertices, a factor of $p^2$ cannot arise explicitly in the
vertex factors, since the only vertex with two factors of momentum in it
is the one-gluon vertex, and in that case the momentum will be $q$. The
only diagram left that has one internal gluon line is Figure
\ref{fig:oneone}b.

The other possibilities have two internal gluon lines, one of which
contracts with an $O_1$ vertex. These are Figures \ref{fig:oneone}a,
\ref{fig:oneone}c, and \ref{fig:oneone}d. When calculating these
diagrams, it is necessary to expand combined gluon denominators as
series in $p^2$, as before.

Figure \ref{fig:oneone}a's symmetry factor is $1 \over 2$, as
for Figure \ref{fig:twotwo}a. The relevant divergent term is
\begin{equation}
        -{5N g^2 \mu^\epsilon \over (4\pi)^2 \epsilon}
        {ig \over 8m^2} T_a p^2 v^\mu.
\end{equation}
Figure \ref{fig:oneone}b gives
\begin{equation}
        +{2 g^2 \mu^\epsilon \over N (4\pi)^2 \epsilon}
        {ig \over 8m^2} T_a p^2 v^\mu
\end{equation}
much in the manner of Figure \ref{fig:twotwo}b's contribution to the
renormalization of $O_2$.  Finally, Figures \ref{fig:oneone}c and
\ref{fig:oneone}d together give
\begin{equation}
        {g^3 \over 8m^2} T_a {5N \over 6} p^2 v^\mu \qint
\end{equation}
or
\begin{equation}
        {5N g^2 \mu^\epsilon \over 3 (4\pi)^2 \epsilon}
        {ig \over 8m^2} T_a p^2 v^\mu.
\end{equation}

\subsection{The $\beta$ function of $O_1$'s coefficient}
The calculation of $\beta_1$ is analogous to that of $\beta_2$.  The
three-point 1PI function of the form of (\ref{eq:o1vertex}), with
class II divergences subtracted, is
\begin{eqnarray}
        \Gamma_{O_1} &=&
        \left\lbrace C_1
        + \left[ \left(6N - {16 \over 3N}\right) C_{kin}^2
        + {10N\over 3} C_{mag}^2
        + \left( -{10N \over 3} + {2 \over N} \right) C_1 \right]
        {g^2 \over (4\pi)^2} \ln \mu \right\rbrace \nonumber \\
        & \times & {g \over 8m^2}
        p^2 v^{\mu} T_a + \cdots
\end{eqnarray}
where the dots represent convergent terms not dependent on $\mu$.
Solving the RGE
\begin{equation}
        \left(  \mu {\partial \over \partial \mu}
             +  \beta_g {\partial \over \partial g}
             +  \beta_{kin} {\partial \over \partial C_{kin}}
             +  \beta_{mag} {\partial \over \partial C_{mag}}
             +  \beta_1 {\partial \over \partial C_1}
             -  2 \gamma_Q - \gamma_A \right) \Gamma_{O_1} = 0
\end{equation}
to order $g^2$ gives
\begin{equation}
        \beta_1 (g(\mu), C_{kin}, C_1(\mu))
        = {g^2 \over (4 \pi)^2} \left[
        \left( -6N+{16\over 3N} \right) C_{kin}^2
        - {10N \over 3} C_{mag}^2
        + {4N \over 3} C_1 \right].
\end{equation}
$C_{kin} = 1$.
At the scale where matching occurs, $C_1 = C_{mag} = 1$, and
\begin{equation}
        \beta_1 (g(m_Q), C_{kin},C_1(m_Q))
        = \left(-8N+{16\over 3N}\right){g^2 \over (4 \pi)^2}.
\end{equation}

Figures \ref{fig:kinkinone}e-f and \ref{fig:oneone}b are the only diagrams
present in an abelian gauge theory. \ref{fig:oneone}b contributes to the
renormalization of $O_1$ in exactly the same way that Figure
\ref{fig:twotwo}b renormalizes $O_2$.  \ref{fig:kinkinone}e-f's
contribution is as above only without the $SU(N)$ factor of $-1 \over
2N$. Therefore, at one loop in a $U(1)$ theory,
\begin{eqnarray}
        \beta_1 &=& {g^2 \over (4 \pi)^2} \left[-{32\over 3}C_{kin}^2
+(4-4)C_1\right] \nl
        &=& -{32\over 3}C_{kin}^2{g^2 \over (4 \pi)^2}
\end{eqnarray}
Other than the gauge coupling constant and the
field normalizations, this is the only thing that runs at one loop to order
$1 \over m^2$ in a $U(1)$ theory.

\section{Reparameterization invariance is satisfied}

In HQEFT, the division of the quark momentum into a large part $mv^\mu$
and a small residual momentum $k^\mu$ is arbitrary, as long as the residual
momentum remains small.  Therefore, there exists a symmetry of
HQEFT called reparameterization invariance, in which the four-velocity
and residual momentum change so as to leave the combination $mv^\mu + k^\mu$
invariant.\cite{luke}  This places constraints on the coefficients of the
correction terms.

However, it is first necessary to define how the heavy quark field
transforms under a reparameterization.  There are at least two
different forms of reparameterization invariance in the literature,
that of Luke and Manohar \cite{luke} and that of Chen \cite{chen}.
Straightforward calculations show that 
the Lagrangian obtained from tree-level matching
(before the field redefinition that removes the class II operators) is
invariant under Chen's transformation, but not under Luke and Manohar's.
Chen's transformation
\begin{eqnarray}
        v^\mu &\rightarrow& v^\mu + \delta v^\mu \nl
        \si &\rightarrow& \left[ 1 + {\delta \slash v \over 2}
        \left(1 + {1 \over 2m + i \vD} \right)\right] \si
        \label{eq:chenfield}
\end{eqnarray}
does the following to the Lagrangian ${\cal L} = \sibar A \si$
to order $1 \over m$:
\begin{equation} \label{deltaL}
        \delta {\cal L} = \sibar \left( \delta A + [ A , im\delta vx]
        + \left \lbrace A, { \delta \slash v \over 2} \right\rbrace
        + {i {\slashD}_\perp \over 4m} \delta \slash v A
        + A \delta \slash v {i {\slashD}_\perp \over 4m} \right)\si
        + O\left( 1 \over m^2 \right)
        \label{eq:chentrans}
\end{equation}
where $\slashD_\perp = \slashD - D \cdot v \slash v$. (Because of the
term proportional to $m$, we would need to know about $1 \over m^3$
terms in the Lagrangian to evaluate what the transformation does to
order $1 \over m^2$.)

Applying (\ref{eq:chentrans}) to (\ref{eq:lagrangian}) induces a change
in the Lagrangian
\begin{eqnarray}
        \delta{\cal L} &=& \sibar (1-C_{kin}) iD \cdot \delta v\si
        +{1 \over 4m} \sibar \left[
        -D \cdot \delta v \vD - \vD D \cdot \delta v \right.
        \nonumber \\
        &&+ \left. (1 - 2C_{mag} + C_2)
        (iD^\mu \delta v^\nu \sigma_{\mu \nu} \vD
        -i\vD D^\mu \delta v^\nu \sigma_{\mu \nu}) \right] \si.
        \label{eq:applied}
\end{eqnarray}

The field-redefinition procedure we have used is equivalent in its
effect on the running of the class I operators to throwing out all
class II terms. Insertions of the class II terms does not affect
the running of the class I terms, because the class I parts
of the loop diagrams involving class II operators are {\it finite.}
The poles that arise from solutions of the classical equation of
motion are eliminated by the vanishing of the class II vertices.
Therefore, our procedure will
yield the same running for the class I operators that we would get
if we kept all of the operators, class I and class II.

The tree-level Lagrangian, including the class II operators, is
symmetric under the RPI symmetry in (\ref{eq:chentrans}). Therefore, even
using our procedure, as long as the regularization
procedure preserves the RPI symmetry in (\ref{eq:chentrans}), the renormalized
Lagrangian to order $1 \over m^2$ must be symmetric under this
transformation, up to terms resulting from the action of the transformation
on the removed class II operators.  The $-D \cdot \delta v \vD
- \vD D \cdot \delta v$ term in (\ref{eq:applied}) results from the
action of the transformation on $O_{D \cdot v}$, so there is no reason
to expect that term to vanish. The others, however, cannot be so
obtained. Therefore, reparameterization invariance sets the constraints
\begin{eqnarray}
        C_{kin} &=& 1 \nl
        2C_{mag} &=& C_2 + 1
\end{eqnarray}
At matching, all of these constants are equal to 1, and the constraints
are satisfied. Under running, in order to maintain these relations,
it must be that
\begin{eqnarray}
        \beta_{kin} &=& 0 \nl
        2 \beta_{mag} &=& \beta_2
\end{eqnarray}
It is well known that the first relation holds to the orders that have
been studied. The second also holds at one loop, according to our results. The
running satisfies reparameterization invariance in the form described by
Chen.

\section{Comparison with other recent calculations}
\label{sec:bal}

In this section, we compare to results in \cite{balzereit,Blo96}

While this work was in preparation, a paper by Balzereit and Ohl
was posted on the net \cite{balzereit}, 
which also calculates the renormalization of the $1/m^2$ operator,
Their technique is quite different from ours; most importantly, they
retain all class II operators, including $O_{D\cdot v}$.  However,
insertions of class II operators do not induce any class I counterterms,
since the amplitudes resulting from class II insertions are finite.
The poles that arise from solutions of the classical equations of 
motion are eliminated by the vanishing of the class II vertices.
Therefore, a calculation using our technique, which assumes
that the Lagrangian contains only class I operators, will yield the
same $\beta$ functions (or, equivalently, anomalous dimensions of local
operators and time-ordered products) obtained in \cite{balzereit} for
the class I operators.

Balzereit and Ohl use a slightly different operator
basis, but their basis for the class I operators is the same as ours up to
class II terms and a sign difference in $O_1$.

Even given the class II terms by which our basis
for the order $1 \over m^2$ local class I operators differs from
\cite{balzereit}, the results still have to be equivalent.  
With both operator bases, our technique can be used
and the calculation will differ in no essential respect. For example,
to renormalize $O_1$ in the basis of \cite{balzereit}, one could look
for terms of the form $2 k' \cdot p v^\mu$, and subtract out class II
terms of the form $(2k' \cdot p - p^2) v^\mu$ by adding twice the
coefficient of $2 p^2 v^\mu$ to the coefficient of $2 k' \cdot p v^\mu$.
This is manifestly the same calculation, and so the results must be the
same, because the class II terms by which our operator bases
differ do not affect the calculated coefficients of the $p^2 v^\mu$ and
$k' \cdot p v^\mu$ terms.

Indeed, our results are equivalent to those of \cite{balzereit}. They
define $O_1$ with the opposite sign, which reverses the signs of the
contributions to $O_1$ from double $O_{kin}$ and double $O_{mag}$
insertions. Our definitions of $\beta$ functions also differ from their
definitions of anomalous dimensions by a further factor of $-4 {g^2/(4
\pi)^2}$. Taking these into account, the terms in our $\beta$ functions
are equivalent to the various elements in their anomalous dimension
matrix ${\hat \gamma}^{(2)}_{phys}$. The nonzero anomalous dimensions in
the first column correspond to the coefficients of $C_1$, $C_{kin}^2$,
and $C_{mag}^2$ in our expression for $\beta_1$. The nonzero anomalous
dimension in the second column corresponds to the coefficient of
$C_{kin} C_{mag}$ in our $\beta_2$.

Very shortly before we posted this paper to the net, another
calculation of the renormalization of the order $1/m^2$ operators
by B. Blok {\em et al.} \cite{Blo96} appeared. This
calculation, like that of Balzereit and Ohl, retaines all
class II operators, and expresses results in the form of
anomalous dimensions mixing local operators with time-ordered
products. The class I operator basis in \cite{Blo96}
is the same as ours up to total derivatives, except for a sign
difference in the definition of $O_{mag}$. The results for the
class I operators also agree with ours, when this sign
difference is taken into account; the first two columns of
the matrix (22a) in \cite{Blo96}
agree with the terms in our beta functions for the case $N=3$.

\section{Conclusions}

We have calculated the running of the heavy quark effective field
theory Lagrangian to order $1 \over m^2$, using a technique in which
continuous field redefinition removes operators from the Lagrangian
which vanish according to the classical equation of motion. Our
results are consistent with symmetry under the reparameterization
transformation of \cite{chen}.
Our results are inconsistent with those of \cite{lee} and
agree with those of \cite{balzereit,Blo96}.

\section*{Acknowledgements}
We thank Howard Georgi for suggesting this topic to us, for important 
discussions, carefully reading the manuscript and checking the calculation.
We also thank Christopher Balzereit and Thorsten Ohl for finding several
errors in an earlier version of this paper.

M.F. would like to thank the members of the Theoretical
Physics Group at Harvard University for their kind hospitality.
This work is supported in part by the National Science Foundation
(Grant \#PHY-9218167) and by the Deutsche Forschungsgemeinschaft.

\appendix
\section{HQEFT Feynman rules to order $1 \over m^2$}

The operators of HQEFT induce various Feynman vertices.  In addition to
the rules listed below, there are of course the usual QCD rules for
gluons; we use background field gauge with a Feynman-like gauge
prescription (at one loop, the gauge fixing parameter does not run and
Feynman-like gauge is OK), so the gluon rules are as given in Abbott
\cite{abbott} with $\alpha = 1$.  The covariant derivative and gluon
field-strength tensor are defined as follows:
\begin{equation}
                D^{\mu} = \partial^{\mu} - igT_a A^{\mu}_a
\end{equation}
\begin{equation}
                G^{\mu \nu} = {i \over g} [D^\mu,D^\nu]
                = \partial^\mu A_a^\nu T_a - \partial^\nu A_a^\mu T_a
                + g f_{abc} T_a A_b^\mu A_c^\nu
\end{equation}
It is convenient to express the rules in terms of
the outgoing heavy quark's residual momentum $k'^\mu$ (taken as
flowing out of the vertex) and the gluon momenta $p^\mu$, $q^\mu$,
and $r^\mu$ (which flow into the vertex). $a$, $b$, and $c$ are
external gluon color indices.  The vertices from the subleading
operators are pictured in Figures \ref{fig:kinmagverts} and \ref{fig:onetwoverts}.

\subsection{Leading-order Lagrangian}
 \subsubsection{propagator}
  \begin{equation}
        {i \over k'\cdot v} \left({1 + \slash v \over 2}\right)
  \end{equation}
 \subsubsection{one-gluon vertex}
  \begin{equation}
        igT_a v^{\mu}
  \end{equation}

\subsection{$O_{kin}$}
 \subsubsection{no-gluon vertex}
  \begin{equation}
        {i \over 2m} k'^2
  \end{equation}
 \subsubsection{one-gluon vertex}
  \begin{equation}
        {ig \over 2m} (2k'-p)^{\mu} T_a
  \end{equation}
 \subsubsection{two-gluon vertex}
  \begin{equation}
        {ig^2 \over 2m} \left \lbrace T_a,T_b \right \rbrace g^{\mu \nu}
         \label{eq:o12g}
  \end{equation}

\subsection{$O_{mag}$}
 \subsubsection{one-gluon vertex}
  \begin{equation}
        {-g \over 2m} \sigma^{\mu \nu} (p_\nu T_a)
  \end{equation}
 \subsubsection{two-gluon vertex}
  \begin{equation}
        {ig^2 \over 2m} \sigma^{\mu \nu} f_{abd} T_d
  \end{equation}

\subsection{$O_1$}
 \subsubsection{one-gluon vertex}
  \begin{equation}
        {-ig \over 8m^2} (p^\mu p \cdot v - p^2 v^{\mu}) T_a
  \end{equation}
 \subsubsection{two-gluon vertex}
  \begin{equation}
        {g^2 \over 8m^2}[ (q-p)\cdot v g^{\mu \nu}
        -(p+2q)^\mu v^\nu + (2p+q)^\nu v^\mu ] f_{abd} T_d
  \end{equation}
 \subsubsection{three-gluon vertex}
  \begin{eqnarray}
        {-ig^3 \over 8m^2} [
        (g^{\mu \rho} v^\nu - g^{\nu \mu} v^\rho)
        f_{bcd}f_{dae}T_e
        &+& (g^{\nu \mu} v^\rho - g^{\rho \nu} v^\mu)
        f_{cad}f_{dbe}T_e \nl
        &+& (g^{\rho \nu} v^\mu - g^{\mu \rho} v^\nu)
        f_{abd}f_{dce}T_e]
  \end{eqnarray}

\subsection{$O_2$}
 \subsubsection{one-gluon vertex}
  \begin{equation}
        {g \over 8m^2} (2k'_\alpha p_\beta \sigma^{\alpha \beta} v^{\mu}
        - p \cdot v (2k'-p)_\alpha \sigma^{\alpha \mu}) T_a
  \end{equation}
 \subsubsection{two-gluon vertex}
  \begin{eqnarray}
                {g^2 \over 8m^2} [ (2k'_\alpha-p_\alpha-q_\alpha)if_{abc}T_c
                (\sigma^{\alpha \mu}v^\nu-\sigma^{\alpha \nu}v^\mu) &&\nl
                + (-p_\alpha\sigma^{\alpha\nu}v^\mu-q_\alpha\sigma^{\alpha\mu}v^\nu
                &+& (p-q)\cdot v \sigma^{\mu\nu})
                \lbrace T^a,T^b \rbrace ]
  \end{eqnarray}
 \subsubsection{three-gluon vertex}
  \begin{eqnarray}
        {-ig^3 \over 8m^2} [
        (\sigma^{\mu \rho} v^\nu + \sigma^{\nu \mu} v^\rho)
        f_{bcd} \lbrace T_a, T_d \rbrace
        &+& (\sigma^{\nu \mu} v^\rho + \sigma^{\rho \nu} v^\mu)
                f_{cad} \lbrace T_b, T_d \rbrace \nl
        &+& (\sigma^{\rho \nu} v^\mu + \sigma^{\mu \rho} v^\nu)
        f_{abd} \lbrace T_c, T_d \rbrace ]
  \end{eqnarray}

\newpage

\begin{figure}
\begin{center}
\mbox{\epsfig{figure=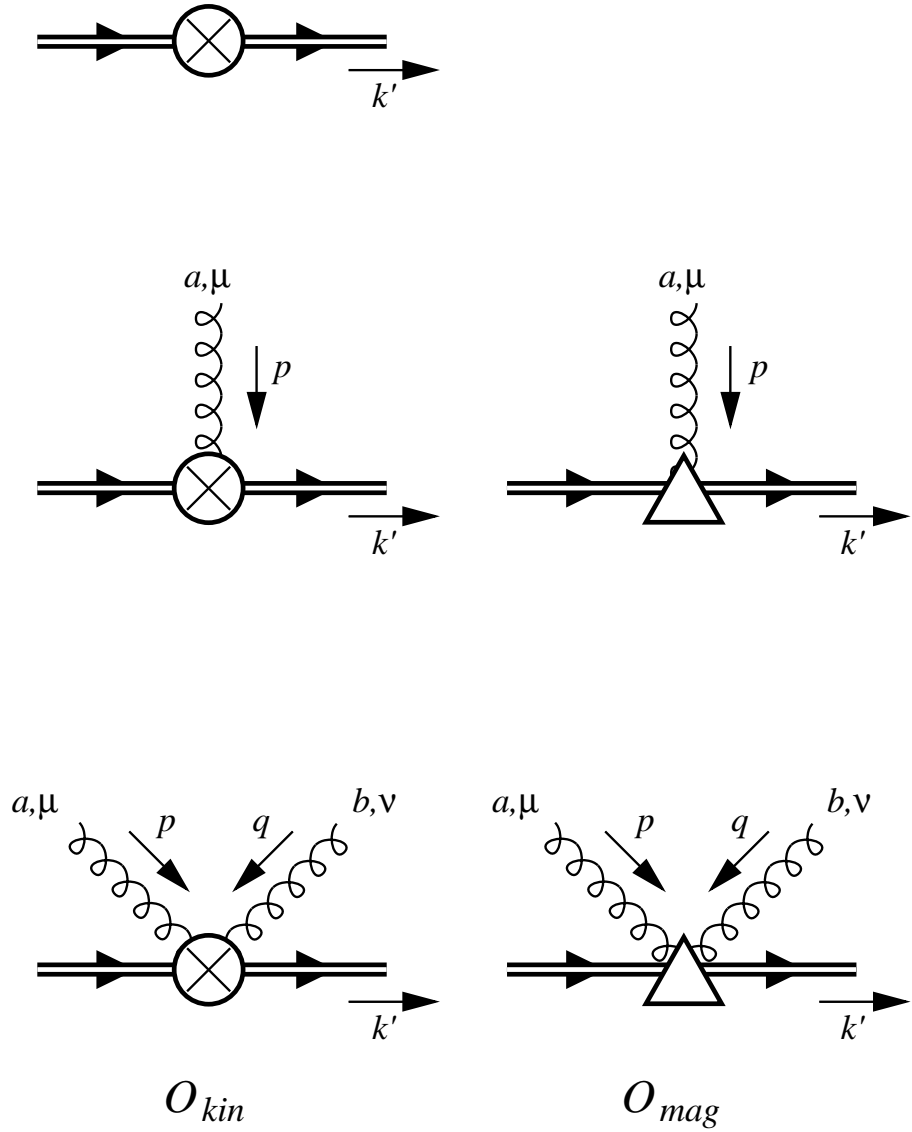,height=15cm}}
\end{center}
\caption{Feynman vertices induced by the
operators $O_{kin}$ and $O_{mag}$.}
\label{fig:kinmagverts}
\end{figure}

\begin{figure}
\begin{center}
\mbox{\epsfig{figure=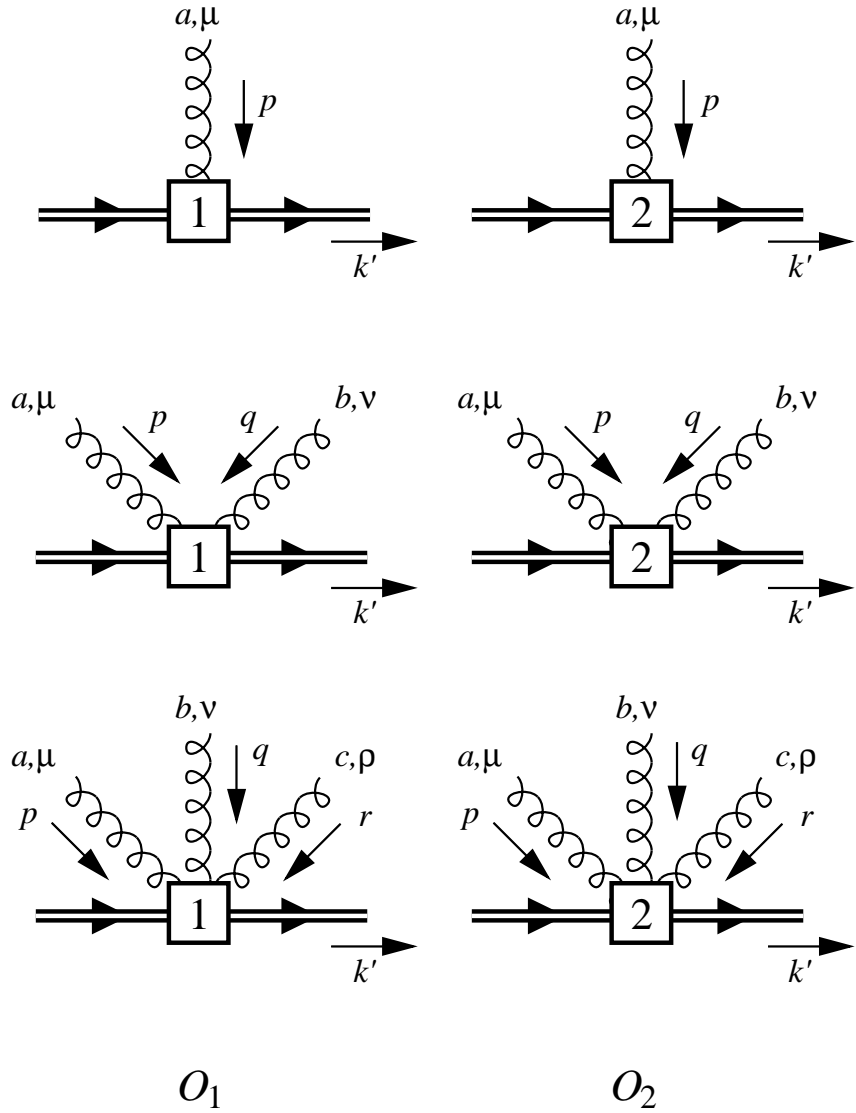,height=15cm}}
\end{center}
\caption{Feynman vertices induced by the
operators $O_{1}$ and $O_{2}$.}
\label{fig:onetwoverts}
\end{figure}

\newpage

\end{document}